# Semantically Configurable Consistency Analysis for Class and Object Diagrams

Shahar Maoz*, Jan Oliver Ringert**, and Bernhard Rumpe

Software Engineering
RWTH Aachen University, Germany
http://www.se-rwth.de/

**Abstract.** Checking consistency between an object diagram (OD) and a class diagram (CD) is an important analysis problem. However, several variations in the semantics of CDs and ODs, as used in different contexts and for different purposes, create a challenge for analysis tools. To address this challenge in this paper we investigate *semantically configurable model analysis*. We formalize the variability in the languages semantics using a feature model: each configuration that the model permits induces a different semantics. Moreover, we develop a parametrized analysis that can be instantiated to comply with every legal configuration of the feature model. Thus, the analysis is semantically configured and its results change according to the semantics induced by the selected feature configuration. The ideas are implemented using a parametrized transformation to Alloy. The work can be viewed as a case study example for a formal and automated approach to handling semantic variability in modeling languages.

> *"One man's constant is another man's variable."*
> Alan Perlis [21]

## 1 Introduction

A class diagram (CD) specifies a model of an object-oriented system structure. The semantics of a CD, that is, its meaning, consists of the (possibly infinite) set of object models it permits. The related kind of diagram, object diagram (OD), is used to document concrete object models. Thus, when both kinds of diagrams are used in a model-driven design process, e.g., when domain experts and engineers use ODs as a means of communication and the latter are responsible for designing the CDs, checking the consistency between a CD and an OD is an important analysis problem. However, several variations and ambiguities in the semantics of CDs and ODs, as they are used in different contexts and for different purposes, create a challenge for analysis tools.

---

* S. Maoz acknowledges support from a postdoctoral Minerva Fellowship, funded by the German Federal Ministry for Education and Research.
** J.O. Ringert is supported by the DFG GK/1298 AlgoSyn.

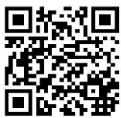

To address this challenge in this paper we investigate *semantically configurable model analysis*. First, we formalize the variability in the semantics of the modeling languages at hand using a feature model: each configuration that the feature model permits, induces a different semantics mapping (over the same domain). Second, we develop a parametrized analysis technique that can be instantiated to comply with every legal configuration of the feature model. Thus, the analysis is semantically configured and its results change according to the semantics induced by the selected feature configuration.

Using a feature model to describe semantic variability has several advantages. First, it provides a means to formally structure the various semantic choices; this supports human comprehension of the semantics, allows comparison of different variants, and, significantly, enables the parsing required in order to support an automatically configurable analysis. Second, the use of a feature model provides a formal means to define logical dependencies between the semantic choices, e.g., mutual exclusion, implication etc. This is indeed necessary, because not all theoretically possible combinations induce sound and useful semantics.

As concrete languages we use the CD and OD sublanguages of UML/P [23]. The semantics of CDs and ODs is based on [5, 7, 10] and is given in terms of sets of objects and relationships between these objects.

Our feature model for the semantics of CD/OD consistency consists of 32 features. One feature, for example, relates to whether empty object models are considered as possible target values in the semantic domain of CDs. Another feature relates to the question of whether incomplete ODs, which describe object models that are missing some attributes or links but can be extended to a complete object model in the semantics of the CD, would be considered consistent with the CD or not. Another feature relates to the semantics of untyped objects in the OD. Each feature is formally defined as part of the CD/OD semantics definition. The feature model organizes the different features so that each of its configurations induces a specific overall semantics.

The consistency analysis itself is realized using a parametrized transformation to an Alloy [13] module. The input for the parametrized transformation consists of a valid configuration of the feature model, a CD, and an OD. The Alloy module is analyzed using a SAT solver and the result shows whether the CD and the OD are consistent given the semantics defined by the configuration. An overview of the architecture of our solution is shown in Fig. 1.

Our work is fully automated and implemented in a prototype Eclipse plug-in, where one can edit CDs and ODs, select a semantic configuration, and check the consistency of a CD and an OD. Feature model definitions and implementation of feature selection use components from FeatureIDE [14]. After the transformation, the Alloy module is analyzed using the APIs of Alloy Analyzer [1].

Sect. 2 discusses related work. Sect. 3 provides a motivating example. Sect. 4 describes the CD and OD languages, their definition of consistency, and the feature models of their semantics. Sect. 5 presents our technique for semantically configurable analysis. Sect. 6 presents the implementation and a discussion. Sect. 7 concludes.

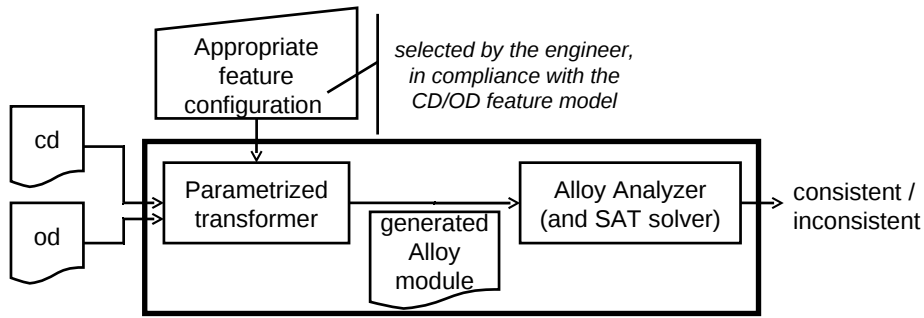

**Fig. 1.** The architecture of our solution

## 2  Related Work

The challenge of semantically configurable analysis has been investigated before in a series of works by Atlee et al. [16, 20, 22, 28], which used *template semantics* to configure the semantics of state machines, and demonstrated configured translations of state machines into SMV and into Java code. Different from these works, we use a feature model to model semantic variability. Moreover, these works relate to state-based behavioral models while our present work focuses on structural models. In this sense, our present work may be viewed as complementary to these previous works.

Previous work in our group [8] has presented a taxonomy of variability mechanisms in language definitions syntax and semantics, and demonstrated the use of feature diagrams to model possible variants. The present work builds on these previous ideas while focusing on semantic variability, specifically, semantic mapping variability (rather than syntactic variability) and on its application to semantically configurable analysis, specifically demonstrated and implemented in the context of CDs and ODs.

Some previous works provide various analyses for CDs (often extended with fragments of OCL), using a translation to a constraint satisfaction problem [6], using ad-hoc algorithms or a direct translation to SAT [12, 26], using a translation to Description Logic [25, 27], or using a translation to Alloy (see, e.g., [2]). We use a transformation to Alloy, but our transformation is very different and much more expressive than the one suggested in [2]. Our transformation extends a basic transformation that we have described in another, more general, paper [17] in two ways: first, it accepts as input not only a CD but also an OD, and second, significantly, it is parametrized based on another input, a feature configuration, so as to support semantically configurable analysis. Finally, to the best of our knowledge, none of the CD analysis works mentioned above support variability-based semantically configurable analysis.

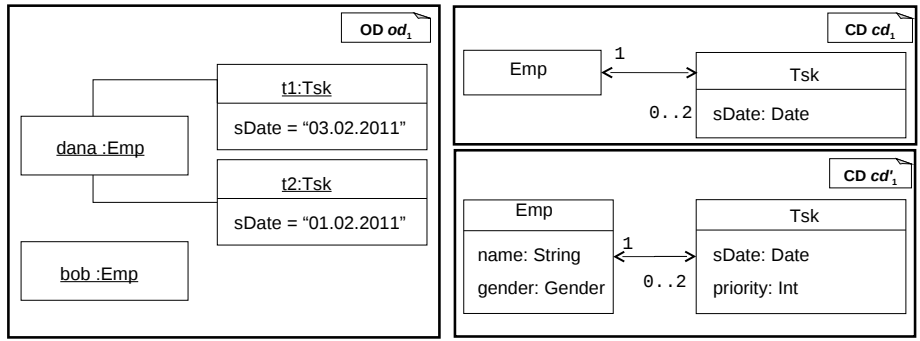

**Fig. 2.** $od_1$, $cd_1$, and $cd'_1$

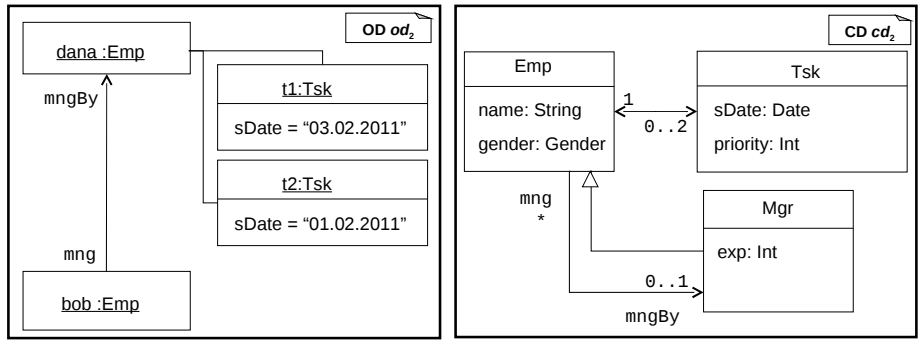

**Fig. 3.** $od_2$ and $cd_2$

## 3 Motivating Example

We describe a simple example to motivate the need for semantically configurable analysis of CD/OD consistency, when CDs and ODs are used for different activities during the development life cycle and in different contexts. The description is semi-formal. Required definitions are given in the following sections.

Consider $od_1$, $cd_1$, and $cd'_1$, shown in Fig. 2. In early stages of system design, a domain expert suggested several ODs as examples of valid system instances, among them $od_1$. $od_1$ consists of employees and tasks: `dana` and `bob` are employees, `dana` has two tasks while `bob` has no tasks. Dana's tasks have a date attribute. The engineers have designed $cd_1$ as a CD for the system and wanted to check the pair $cd_1$/ $od_1$ for consistency before they continue.

Later in the design process, after more requirements elicitation, additional information became available and the CD $cd_1$ evolved into a more detailed one, $cd'_1$, where the same classes include additional attributes. The engineers wanted to check the consistency of $cd'_1$ not only against some new ODs, where all the new attributes are defined, but also against the older OD $od_1$, which includes only a partial list of attributes. Although the objects in $od_1$ did not include all the attributes shown in the new CD, the engineers expected that $od_1$ would be considered consistent with $cd'_1$, because it could be extended into a complete valid instance of $cd'_1$ where more attributes are present.

After a design review, another version of the CD was prepared, $cd_2$, as shown in Fig. 3. In $cd_2$ a new class Manager was added as a specialization of Emp, and a related association with roles mngBy and mngs. In turn, the domain expert used $od_1$ to create $od_2$, by adding a link between bob and dana, so as to specify that dana manages bob. While in $od_2$ dana's shown type is Emp, it is understood that dana is also a Manager, because she manages bob. The engineers wanted to check the consistency of $cd_2$ and $od_2$ and expected the result to be positive.

A test engineer, responsible for creating test cases that will be executed after a running prototype of the system is created, wanted to specify each test's pre- and post-conditions using ODs. As a sanity check, it was necessary to verify the consistency between each of these ODs and the system's CD. In this case, a much more strict and complete semantics was assumed, i.e., that the instances in the OD include complete lists of attributes and specify their exact type, otherwise the tests may not be accurate or fail (e.g., if dana is constructed as an employee rather than as a manager). Thus, to be useful, CD/OD analysis in the context of testing required a slightly different semantics. Note that based on this semantics, $cd_2$ and $od_2$, which have been considered consistent in the context of requirement elicitation, are not considered consistent anymore.

Moreover, the design team noted that the objects in the system may be dynamically constructed and destructed: the system starts with no object instances, and during execution may return to this "no instance" state. Thus, an empty OD, representing the empty OM, should be considered a valid system instance, because, for example, it needs to be used as a pre- or post-condition of some tests. Therefore, despite common standard definitions elsewhere and perhaps against many modelers' intuition, when checking this empty OD against the system's CD for consistency, the team expected a positive result.

Finally, the most complete and detailed version of the system's CD (not shown here) is intended for skeleton code generation of the actual implementation. While in this CD no classes or attributes may be omitted, the team wanted to check it against all ODs used in the design and see that they are consistent.

This example demonstrates that the consistency of a given CD and OD depends on the specific usage of the diagrams and the context in which the question arises; it thus shows the need for more than one definition of semantics for CD/OD consistency. Characterizing and formalizing the required variability, and showing how it can be implemented in a single, configurable analysis solution, are the challenges we address in this paper.

## 4  CDs and ODs, Consistency, and Semantic Variability

### 4.1  Class and object diagrams languages

The concrete CD and OD languages we use are sublanguages of UML/P [23], a conceptually refined and simplified variant of UML designed for low-level design and implementation. Our semantics of CDs is based on [5, 7, 10] and is given in terms of sets of objects and relationships between these objects. More formally, the semantics is defined using three parts: a precise definition of the syntactic domain, i.e., the syntax of the modeling language CD and its context conditions (we use MontiCore [15, 19] for this); a semantic domain, for us, a subset of the System Model (see [5, 7]) OM, consisting of all finite object models; and a mapping $sem : CD \to \mathcal{P}(OM)$, which relates each syntactically well-formed CD to a set of constructs in the semantic domain OM. The semantics of ODs is defined over the same semantic domain OM, using a mapping $sem : OD \to \mathcal{P}(OM)$, which relates each syntactically well-formed OD to a set of constructs in the semantic domain OM, that is, to a set of object models. Note that the semantic domain of CDs is made of OMs, not ODs. For a thorough and formal account of the semantics see [7].

For example, the semantics of $cd_1$ shown in Fig. 2 includes all object models consisting of tasks and employees where each employee is responsible for up to two tasks, and each task is done by exactly one employee and has an attribute `sDate` of type `date`. Note that the empty object model, which is an object model with no objects at all, may or may not be considered in the semantics of this CD. In addition, note that we did not say whether object models whose tasks have additional attributes may be considered in the semantics of this CD or not. As another example, the semantics of $od_1$ shown in Fig. 2 includes all object models consisting of two employees where one of the employees is linked to two tasks that have certain `sDate` values. Note that we did not say whether object models that have additional employees, with or without tasks, should be considered in the semantics of this OD or not. These ambiguities and possible variations are examples of the kinds of semantic variability that affect the CD/OD consistency check, as we discuss below.

Finally, we support the following CD language constructs: class attributes, enumerations, uni- and bi-directional associations with multiplicities, aggregation, composition, generalization (inheritance), interface implementation, and abstract and singleton classes. The OD language constructs we support include objects, their attributes, and the links between them.

### 4.2 Consistency

A set of diagrams is considered *consistent* if the intersection of the semantics of all diagrams in the set is not empty [4]. Formally:

**Definition 1 (consistency).** *Given a set of diagrams D, we say that D is consistent iff* $\bigcap_{d \in D} sem(d) \neq \emptyset$.

By applying the above definition to the special case of a CD and an OD we get:

**Definition 2 (CD/OD consistency).** *Given a CD cd and an OD od, we say that the cd and od are consistent iff* $sem(cd) \cap sem(od) \neq \emptyset$.

While the definition of consistency is generally accepted, definitions of the semantic mapping function *sem*, for CDs and ODs, may vary. To formally handle variability in the semantics mapping we use the feature models described next.

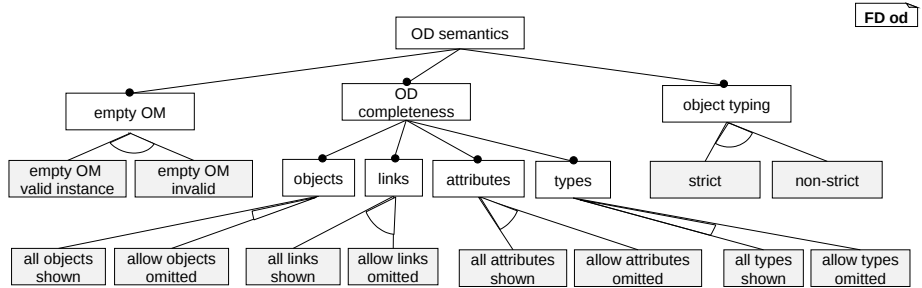

**Fig. 4.** The OD semantics feature diagram

### 4.3 The semantic variability feature models

A feature model describes a structured set of features and their logical dependencies [3, 9]. Feature models are commonly used in the area of software product lines. They may be visually represented using feature diagrams, which are basically and-or trees, extended with textual cross-tree logical constraints. Here we use a feature model to formalize variability in the semantics of CDs and ODs. The model is composed of two sub-models, for CD semantics and for OD semantics, and of several cross-tree logical constraints. In the diagrams we use the standard notation: for mandatory features, a line ending with a filled circle; for alternative features of which exactly one must be selected (xor), an empty slice covering the lines leading to the different alternatives.

Our feature model for OD semantics consists of 19 features, as shown in the feature diagram in Fig. 4. Roughly, a valid feature configuration of this model specifies whether the empty object model may be considered a valid OM, whether the objects shown, links shown, attributes shown, and types shown are complete or not, and whether all objects shown in the diagram must be typed with their most specific type, or can use one of their super types.

Our feature model for CD semantics for CD/OD consistency contains 11 features, as shown in the feature diagram in Fig. 5. A valid feature configuration of this model specifies whether the empty object model may be considered a valid instance of a CD, whether the lists of attributes shown are considered complete or not, and whether the set of classes shown is considered complete or not.

The complete feature diagram for CD/OD consistency feature model is built from a CD/OD consistency feature at the root, using the two feature diagrams described above to represent required features, as its sub trees, as shown in Fig. 6. To this composed diagram we add cross-tree logical constraints that define dependencies between the different features, for us, the semantic choices, e.g., mutual exclusion, implication etc. This is indeed necessary, because, as we have found also during evaluation (see Sect. 6), not all theoretically possible combinations (feature configurations) induce sound and useful semantics. Specifically, we add the following 3 constraints:

$$\textbf{not } (\ cd.allowClassesOmitted \textbf{ and } od.allowTypesOmitted\ ) \qquad (1)$$

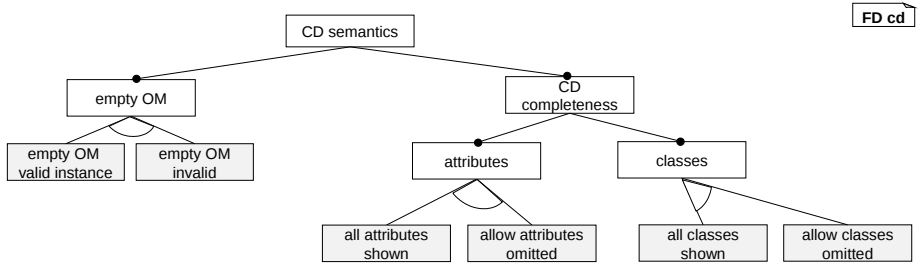

**Fig. 5.** The CD semantics feature diagram

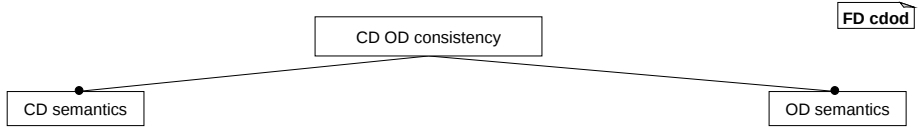

**Fig. 6.** The composed CD/OD semantics feature diagram

$$od.allowObjectsOmitted \textbf{ implies } od.allowLinksOmitted \quad (2)$$
$$cd.emptyOMInvalid \textbf{ iff } od.emptyOMInvalid \quad (3)$$

We add constraint 1 because the combination of allowing classes to be omitted from the CD (which means allowing instances to include objects of classes not shown in the CD) and of allowing the OD to include untyped objects, results in a semantics which is much too permissive and is not useful. We add constraint 2 because if objects are allowed to be omitted, the links they could have been connected with must also be allowed to be omitted. We add constraint 3 because having the empty OM in the semantics of CDs while excluding it from the semantics of ODs (or vice versa) does not make sense.

Overall, our feature model contains 32 features, 14 of which are core features, which are included in all configurations. The model has 144 valid configurations. The complete feature model used in our work is available in [24], also in a format compliant with [18], to allow others to inspect it and use it.

## 5 Semantically Configurable Consistency Analysis

The key to the semantically configurable consistency analysis is a parametrized transformation to an Alloy [13] module. In addition to a CD and an OD, the input for the parametrized transformation includes one valid configuration of the CD/OD consistency feature model described in the previous section.

We now describe the parametrized transformation to Alloy. A variant of our transformation, which takes only a CD as input and is not semantically configurable, is presented in [17]. Here we give an overview of the generated Alloy module and then focus on the parts related to handling variability. We use the CDs and ODs presented earlier in Sect. 3 as running examples.

## 5.1 Overview of the transformation to Alloy

The basic transformation relies on several foundational signatures and facts. These include an abstract signature `FName`, used to represent association role names and attribute names for all classes; an abstract signature `Obj`, which serves as the parent of all classes, and whose `get` Alloy field relates it and an `FName` to instances of `Obj` (this allows more flexibility than the built-in Alloy fields); an abstract signature `Val` as a specialization of `Obj`, used to represent all predefined types (i.e., primitive types and other types that are not defined as classes in the CD); a signature `EnumVal`, which extends `Obj` too, and is used to represent values of enumeration types; and several facts, among them ones that state that enumeration values as well as primitive values can have no further fields and should only appear in an instance if referenced by an object.

A number of parametric predicates are used to specify constraints such as association's multiplicities and directions. These are instantiated with concrete values from within the CD predicate described next. Rather than using Alloy's `extends` keyword to specify generalization relations, we use generated functions that return the set of sub classes of each class, e.g., if `Mgr` is a specialization of `Emp` then the function `EmpSubs` returns the atoms in {`Emp`, `Mgr`}.

The CD and the OD themselves are represented using two predicates, `pred cd` and `pred od`. In `pred cd` the attributes and associations of each class are defined and then restricted using the multiplicity and directionality predicates mentioned above. In `pred od` the existence of the objects is stated and their attributes and links are defined.

Finally, a predicate `pred consistentCDOD` is defined, consisting of the single statement `cd and od`. Checking consistency is done by executing Alloy Analyzer `run` command for `consistentCDOD`.

## 5.2 Handling semantic variability

Handling variability is technically realized using generated parametrized Alloy predicates and their instantiation from within `pred cd` and `pred od`. Below we show how some of the features are handled.

**OD features.** List. 1.1 shows several parametrized Alloy predicates corresponding to the different features available for OD semantics. As a concrete example, List. 1.2 shows the predicate that represents an OD, specifically $od_2$, presented earlier in Sect. 3, Fig. 3, in the context of a specific semantic configuration where the empty OM is not a valid instance, all objects and links are shown but attributes may be omitted, all types are shown but are not strict. We now explain the two listings in detail.

First, the predicate `emptyOMNotValidOD` (List. 1.1 line 2) specifies that there exists at least one object. It is mentioned in `pred od` iff the semantic configuration includes the feature *od.emptyOMInvalid* (see List. 1.2 line 18).

Second, the predicates in lines 5-16 are used to specify the three completeness features, for objects, links, and attributes. The predicate `allObjectsShownOD`

```
1  // Semantic variation feature: empty OM
2  pred emptyOMNotValidOD { some Obj }
3
4  // Semantic variation feature: OD completeness
5  pred allObjectsShownOD[objs: set univ] {
6    univ = (objs + FName + auxilary + Val + EnumVal + Int) }
7
8  pred allLinksShownOD[obj: Obj, roleNames: set FName] {
9    no {obj.get[FName - roleNames] - Val - EnumVal } }
10 pred allLinksShownODCmplt[obj: Obj, roleName: one FName,
11   partners: set Obj] { obj.get[roleName] = partners }
12 pred allLinksShownODIncmplt[obj: Obj, roleName: one FName,
13   partners: set Obj] { partners in obj.get[roleName] }
14
15 pred allAttribShownOD[obj: Obj, definedAttrs: set FName] {
16   obj.get.(Val + EnumVal) = definedAttrs }
17
18 // Semantic variation feature: object typing
19 pred strictTypingOD[obj: univ, type: set univ] {
20   obj in type }
21
22 pred nonStrictTypingOD[obj: univ, subtypes: set univ] {
23   obj in subtypes }
```

**Listing 1.1.** Parametrized Alloy predicates for OD semantics features

specifies that the set of objects it receives in its parameter (plus some other atoms from utility sets used in our translation) is equal to the module's universe, i.e., that there are no more objects except the ones specified in its parameter. An example instantiation of this predicate appears in line 16 of List. 1.2, specifying that dana, bob, and the two tasks, as shown in the diagram, are all the objects in the object model. The other completeness predicates use the get relation defined in our translation; this special representation of object's attributes and field names allows us to specify their presence or absence. In our example, the semantic configuration requires that all links are shown, and so lines 8-15 of List. 1.2 instantiate the allLinksShownOD and the allLinksShownODCmplt predicates for all the links in $od_2$.

Third, the last two predicates in List. 1.1 handle strict and non-strict typing: both specify that the set of objects in the first parameter is included in the set of objects assigned to the second parameter. We keep the two predicates separate for better readability when they are used: strict typing is used with a specific signature while non-strict typing is used with our translation's sub classes functions (see above). In our example we chose non-strict typing so we use the sub classes functions EmpSubs, which returns the atoms in {Emp, Mgr}, and TskSubs, which returns the atoms in {Tsk} (lines 5-6 of List. 1.2).

```
1 pred od2 {
2   some dana: Obj| some bob: Obj| some t1: Obj| some t2: Obj|
3   # {dana + bob + t1 + t2} = 4
4   // Semantic variation feature: object typing
5   and nonStrictTypingOD[dana + bob, EmpSubs]
6   and nonStrictTypingOD[t1 + t2, TskSubs]
7   // Semantic variation feature: OD completeness
8   and allLinksShownOD[dana, worksOn]
9   and allLinksShownOD[bob, mngBy]
10  and allLinksShownOD[t1, doneBy]
11  and allLinksShownOD[t2, doneBy]
12  and allLinksShownODCmplt[bob, mngBy, dana]
13  and allLinksShownODCmplt[dana, worksOn, {t1 + t2}]
14  and allLinksShownODCmplt[t1, doneBy, dana]
15  and allLinksShownODCmplt[t2, doneBy, dana]
16  and allObjectsShownOD[dana + bob + t1 + t2]
17  // Semantic variation feature: empty OM
18  and emptyOMNotValidOD }
```

**Listing 1.2.** Example Alloy predicate for $od_2$ (shown in Fig. 3)

**CD features.** List. 1.3 shows the parametrized Alloy predicates related to the different features available for CD semantics. As a concrete example, List. 1.4 shows the predicate that represents a CD, specifically $cd_2$, presented earlier in Sect. 3, Fig. 3, in the context of a specific semantic configuration where the empty OM is not part of the semantics, all classes are shown and their list of attributes is complete. We now explain the two listings in detail.

The predicate emptyOMNotValidCD (List. 1.3 line 2) specifies that there exists at least one object. It is mentioned in pred cd iff the semantic configuration includes the feature $cd.emptyOMinvalid$ (just like in pred od).

The remaining predicates in List. 1.3 handle completeness. The predicate allAttribShownCD specifies that the get relation of the object does not include any field name outside the set of field names specified in the fNames parameter (see List. 1.4 lines 17-19 for instantiations with all classes and their field names). The predicate allowMoreAttribCD specifies that for the signature given as the objs parameter either there are no more fields than specified in the fNames parameter or there are additional attributes and enumeration values. The predicate allClassesShownCD specifies that the model's universe will only contain object instances of the classes given as a parameter. It is instantiated in List. 1.4 line 20 with all classes shown in $cd_2$.

## 6 Implementation and Discussion

**Implementation.** We have created a prototype implementation of our work, packaged as an Eclipse plug-in. For the representation of the CD/OD semantics feature model and the selection of valid configurations we use components from

```
1  // Semantic variation feature: empty OM
2  pred emptyOMNotValidCD { some Obj }
3
4  // Semantic variation feature: CD completeness
5  pred allAttribShownCD[objs: set Obj, fNames:set FName] {
6    no objs.get[FName - fNames] }
7
8  pred allowMoreAttribCD[objs: set Obj, fNames:set FName] {
9    all f : (FName - fNames) | (
10   (no objs.get[f])
11   or (one v : Val | all o : objs |  o.get[f] = v)
12   or attribOfEnumValue[objs, f]  ) }
13
14 pred allClassesShownCD[objs: set Obj] {
15   univ = (objs + FName + auxilary + Val + EnumVal + Int) }
```
**Listing 1.3.** Parametrized Alloy predicates for CD semantics features

FeatureIDE [14]. For editing CDs and ODs we use parsers and editors (with syntax highlighting etc.) generated by MontiCore [15, 19]. The transformation to Alloy uses FreeMarker templates [11]. Analysis is done using Alloy's APIs [1]. The prototype plug-in together with several examples is available from [24].

**On semantic variability.** One may consider semantic variability in a modeling language definition to be a weakness, as it may create confusion and lead to ambiguities in its comprehension and use. We believe, however, that for general purpose languages such as the sub-languages of the UML, state machines, class diagrams, etc., a certain degree of variability in general, and of semantic variability in particular, is a necessity. The very 'general purpose' nature of the language dictates that it will be used for a variety of tasks and in different contexts, which, in practice, entails a requirement for variability. This is evident also from the works of Atlee et al. [20, 22, 28]. Still, we do not try to promote the existence of too many semantics; instead, we aim to formally and precisely define the specific points where the semantics should vary and automate the application and use of the possible resulting definitions.

As an alternative to language level semantic variability, one may suggest to enrich the language syntax with keywords that allow the modeler to explicitly choose between variants, e.g., by adding optional keywords such as 'complete' / 'incomplete', 'strict' / 'permissive' etc. as modifiers, at the diagram level or the diagram-element level. The advantage of this is that there is a single semantics to handle. The disadvantages however are (1) that the language syntax becomes more complicated, (2) that questions may arise regarding the default semantics, e.g., if the 'complete' /'incomplete' keywords are omitted, and, significantly, (3) that this solution does not support cases where the same diagram should change its meaning in different phases of the development process (e.g., when the same CD should be considered complete during design but incomplete during

```
1  pred cd2 {
2    // Definition of class attributes
3    ObjAttrib[Tsk, priority, type_Int]
4    ObjAttrib[Tsk, sDate, type_Date]
5    ObjAttrib[Emp, gender, GenderEnum]
6    ObjAttrib[Emp, name, type_String]
7    ObjAttrib[Mgr, gender, GenderEnum]
8    ObjAttrib[Mgr, exp, type_Int]
9    ObjAttrib[Mgr, name, type_String]
10   // Associations
11   ObjLUAttrib[EmpSubs, mngBy, MgrSubs, 0, 1]
12   ObjL[MgrSubs, mngBy, EmpSubs, 0]
13   BidiAssoc[EmpSubs, worksOn, TskSubs, doneBy]
14   ObjLUAttrib[TskSubs, doneBy, EmpSubs, 1, 1]
15   ObjLUAttrib[EmpSubs, worksOn, TskSubs, 0, 2]
16   // Semantic variation feature: cd completeness
17   allAttribShownCD[Tsk, priority+sDate+doneBy]
18   allAttribShownCD[Emp, gender+name+mngBy+worksOn]
19   allAttribShownCD[Mgr, gender+exp+name+mngBy+worksOn]
20   allClassesShownCD[Tsk+Emp+Mgr]
21   // Semantic variation feature: empty OM
22   emptyOMNotValidCD }
```

**Listing 1.4.** Example Alloy predicate for $cd_2$ (shown in Fig. 3)

analysis). It is important to note, though, that our work can easily be adapted to support this solution: the only change is that the 'configuration' would not come from the feature model but from the keywords on the diagrams themselves.

**Evaluation of our solution.** Our choice of Alloy as the target formalism for analysis was motivated by Alloy's expressive power, its readability, and its readily available automated analysis. Still, it is important to note that Alloy's analysis is generally bounded by a user-defined scope. Interestingly, however, in the context of CD/OD consistency, the scope limitation is relevant to some semantic configurations but is irrelevant to others: specifically, when the CD and OD semantics assume that the diagrams show all classes and all objects, the scope to be used can be calculated from the input and the analysis is sound and complete. That said, our experience with Alloy shows that it does not scale well for large scopes. Alloy was not designed to scale, see the small scope hypothesis discussed in [13].

We have validated our work as follows. First, we created an automated test that generates all 144 legal configurations of our feature model, checks their application to the consistency check of three different CD OD pairs, and verifies that the result is correct. Second, we used FeatureIDE's user interface to manually define 7 different configurations, we used MontiCore's generated CD and OD editors to edit 12 CDs and ODs (including the ones shown in this paper in Sect. 3), we ran the configurable consistency check using our plug-in and ob-served that the results are correct. Moreover, we have pre-prepared a number of configurations that we believe are most useful for specific task contexts, e.g., for requirements elicitation and for testing. All configurations, CDs, and ODs used in our validation are available with the implemented plug-in from [24]. We encourage the interested reader to check them.

One lesson learned during evaluation was the importance of constraints between features (the second constraint presented in Sect. 4.3, relating object omission with links omission, was discovered in the course of our experiments). Another lesson learned relates to scalability. While our implementation works very fast for small CDs and ODs, it does not scale to handle CDs associations with high multiplicities and ODs with many objects. As mentioned above in the discussion of the use of Alloy, scalability will require the use of abstractions or the development of a different analysis approach. Finally, one may suggest additional CD features we do not yet support (e.g., constrained generalization sets, a fragment of OCL constraints etc.) and additional semantic variation features (e.g., allow role names omitted in the OD). Our work can be extended to support these additions. Each additional feature will require corresponding support in the configurable transformation and possibly logical constraints on its combination with other features. We leave these for future work.

## 7 Conclusion

In this paper we have investigated the idea of semantically configurable analysis in the context of CD and OD consistency. We formalized semantic variability in these languages using a feature model and presented a semantically configurable fully automated analysis solution based on a parametrized transformation to an Alloy module and its analysis with a SAT solver. The work was implemented in an Eclipse plug-in and demonstrated with examples.

We consider the following possible future work. First, extending our work to support additional CD language features, e.g., constrained generalization sets. Second, defining feature models for semantic variability in other modeling languages and developing related parametric analysis problems, e.g., the model-checking of a statechart against a sequence diagram.